\documentclass[12pt]{article}
\usepackage{epsfig}
\usepackage{graphicx}

\def\beq{\begin{equation}}
\def\eeq#1{\label{#1}\end{equation}}
\def\eeqn{\end{equation}}

\def\beqa{\begin{eqnarray}}
\def\eeqa#1{\label{#1}\end{eqnarray}}
\def\eeqan{\end{eqnarray}}

\let\bar=\overbar

\def\Dslash{\not{\hbox{\kern-4pt $D$}}}
\def\dslash{\not{\hbox{\kern-2pt $\del$}}}

\def\msb{{\bar{\ssstyle M \kern -1pt S}}}

\usepackage{fancyhdr,graphicx}
\def\Title#1{\begin{center} {\Large {\bf #1} } \end{center}}

\begin{document}

\Title{An internal mechanism for the anti-glitch observed in \mbox{AXP 1E 2259+586}}

\bigskip\bigskip


\begin{raggedright}

{\it 
Ignacio F. Ranea-Sandoval$^{1}$~~and Federico Garc\'{\i}a$^{2}$\\
\bigskip
$^{1}$Grupo de Gravitaci\'on, Astrof\'{\i}sica y Cosmolog\'{\i}a, Facultad de Ciencias Astron\'omicas y Geof\'{\i}sicas, Universidad Nacional de La Plata.\\ Paseo del Bosque S/N 1900. La Plata, Argentina\\
\bigskip
$^{2}$Instituto Argentino de Radioastronom\'{\i}a, CCT La Plata - CONICET, C.C. 5 (1984) Villa Elisa, Buenos Aires, Argentina \\ Facultad de Ciencias Astron\'omicas y Geof\'{\i}sicas, Universidad Nacional de La Plata. Paseo del Bosque S/N 1900. La Plata, Argentina. 
}

\end{raggedright}

\section{Introduction}

Magnetars are fascinating objects that are thought to be neutron stars powered by their strong internal magnetic fields \cite{duncan1992}. 

There is evidence that neutron stars suffer a long term spin-down. Moreover, many sudden spin-ups, known as glitches in the literature, have been observed in pulsars and magnetars \cite{glitches1,glitches2}. 

Clear evidence of a sudden spin-down was detected in the Anomalous X-ray Pulsar \mbox{AXP 1E 2259+586}, an object cataloged as a magnetar \cite{antiglinch}. This event received the name ``anti-glitch''. 
To adjust the timing data from {\it Swift}, two different interpretations for the observational evidence were proposed: (i) an anti-glitch in which $\Delta \nu / \nu = -3.1(4) \times 10^{-7}$ followed by a spin-up event of amplitude $\Delta \nu / \nu = 2.6(5) \times 10^{-7}$; (ii) an anti-glitch in which $\Delta \nu / \nu = -6.3(7) \times 10^{-7}$ followed by a second anti-glitch in which $\Delta \nu / \nu = -4.8(5) \times 10^{-7}$ \cite{antiglinch}. Based on a bayesian analysis, model (ii) is favored \cite{hu2013}.

Regarding the energetics related with this event we mention that consistently with the epoch of the anti-glitch, {\it Fermi}/GBM detected a hard X-ray burst with a duration of 36 ms \cite{foley.et.al}. The observed fluence in the 10--1000~keV band corresponds to an energy release of $E_{\gamma} \sim 10^{38}$~erg. Moreover, an increase by a factor 2 in the 2--10~keV flux was also observed \cite{antiglinch}, resulting in a $E_{\rm X} \sim 10^{41}$~erg energy release \cite{antiglitch-exp1}.

\mbox{AXP 1E 2259+586} has a characteristic age of $\sim$10$^6$~yr, a $\sim$7~s period and a spin-inferred surface dipolar magnetic field of $B_{d} \sim 5.9 \times 10^{13}$~G\footnote{http://www.physics.mcgill.ca/$\sim$pulsar/magnetar/main.html}. Over the last two decades it has been monitored by the {\it Rossi X-ray Timing Explorer} and the {\it Swift X-ray Telescope}. With the exception of two spin-up glitches in 2002 \cite{kaspi2003} and 2007 \cite{idem2012}, a timing event in 2009 \cite{idem2012} and this anti-glitch in 2012, the source showed a stable spin-down rate.

Several explanations for this anti-glitch event have been proposed, which can be divided in two different families of models: one based on an external \cite{antiglitch-exp1,antiglitch-exp2,antiglitch-exp3,antiglitch-exp4} origin and the other on an internal \cite{duncan2013,garran2015} one. Despite several searches in radio and X-ray wavelengths, no surrounding afterglow was detected \cite{antiglinch}, arguing against a sudden particle outflow or wind-driven scenario. In this sense, a more promising approach in order to explain the phenomenum seems to be an internal rearrangement of the star. 

In this work we present a simple internal mechanism which could account for the observed sudden spin-down of the star \cite{garran2015}. The central idea behind this model is that as a consequence of the natural long term decay of the internal magnetic field, an initially prolate-shaped stable stellar configuration becomes unstable enough to crack the crystallized stellar crust. Then, the re-accommodation of the star into a stable more-spherical shape, could naturally lead to the occurrence of an anti-glitch, as a consequence of the conservation of angular momentum. A similar scenario was also suggested to account for the \mbox{SGR 1900+14} event \cite{ioka}.

\section{The proposed mechanism}
Deformations for a rotating uniform-density with a mixed poloidal-toroidal magnetic field configuration were calculated in \cite{cutler.2002,mag-deform}. Studies with a wider and more realistic family of equations of state were performed recently by \cite{frieben2012}. As in both set of works, the quadrupolar distortions obtained are of the same order of magnitude, in our model, we used a simple uniform density star which allows to perform analytical calculations. In addition, since magnetars like \mbox{AXP 1E 2259+586} show long spin periods, deformations due to rotation are negligible, and thus, we do not consider them on our model. All these different works conclude that equilibrium configurations for stars with strong poloidal magnetic fields or in rapid rotation, are oblate while, when internal toroidal fields dominate the magnetic field configuration, prolate stars result favored.

For incompressible stars of uniform density and mixed poloidal-toroidal magnetic field, the quadrupolar distortion of equilibrium configurations, $\epsilon$, of the $l=2$ volume preserving mode is given by:

\begin{equation} \label{epsilon}
\epsilon = \frac{I_{ zz} - I_{xx}}{I_{zz}} = - \frac{25R^4}{24G_NM^2}\left(\langle B_t^2\rangle-\frac{21}{10}\langle B_p^2\rangle\right),
\end{equation}

\noindent where $M$ and $R$ are the mass and the radius of the undeformed star, respectively. $G_N$ is the gravitational constant and $\langle B_{t,p}^2\rangle$ is the mean value of the square of the toroidal and poloidal magnetic field strengths.

A purely toroidal magnetic field configuration is known to be unstable \cite{stable}, but a poloidal component with energy $E_p/E_t = B_p^2/B_t^2 \sim 1-5\%$ can stabilize the magnetic field configuration \cite{reisenegger2013}. Thus, we neglect the poloidal contribution to equation (\ref{epsilon}) considering that $\langle B_p^2\rangle \ll \langle B_t^2\rangle$.

Neutron star crusts are strong enough to support ellipticities up to a critical value of $\epsilon _c \le  4 \times 10^{-6}$ before cracking \cite{NScrust}. 


After these general considerations we present the theoretical picture that want to explore: given an ``initial'', mostly toroidal, magnetic field with strength $\langle B_t^i\rangle$, the neutron star crust crystallizes in a prolate equilibrium configuration, with an ellipticity $\epsilon ^{-}$ given by equation~(\ref{epsilon}). A series of effects that are thought to take place inside the neutron star produce a progressive decay of the original magnetic field in time scales $\sim 10^{5}$~yr \cite{pons2009,vigano2013}. As a consequence, the prolate configuration with $\epsilon ^-$ departs from equilibrium until the stellar crust reaches a critical strain and cracks. Then, the stellar structure achieves a new stable and less prolate configuration, with ellipticity $\epsilon ^{+}$, associated to the present or ``final'' magnetic field strength, $\langle B_t^f\rangle$. Because the more spherical configuration, $\epsilon^+$, has greater moment of inertia respect to the spin axis, chosen to be $z$ in our case, with respect to the previous $\epsilon^-$ configuration, and considering that in the absence of an external torque angular momentum conserves, this sudden change in the stellar structure can easily account for the observed sudden frequency spin-down.
 
A change in the oblateness of a uniform density star induces a spin frequency shift given by





\begin{equation} \label{plot}
\frac{\Delta \nu}{\nu} = \frac{(1-2\epsilon ^+)^{1/3}}{(1-2\epsilon ^-)^{1/3}} - 1 \approx \frac{2}{3}(\epsilon^- - \epsilon^+),
\end{equation}



\noindent relationship that, as a function of $\langle B_t^{i,f} \rangle ^2$, can be written as (for details on the calculations see \cite{garran2015}) 

\begin{equation} \label{deltanu1}
\frac{\Delta \nu}{\nu} = \frac{2}{3} \frac{25 R^4}{24 G_N M^2} \left( \langle B_t^{f}\rangle ^2- \langle B_t^{i}\rangle ^2\right),
\end{equation}

\noindent from where it follows that 

\begin{equation} \label{deltanu2}
\langle B_t^{f} \rangle \approx \sqrt{ \langle B_t^{i} \rangle ^2- B_0^2 }.
\end{equation}

\noindent where $B_0^2 = - \frac{3}{2} \frac{\Delta \nu}{\nu} \frac{24 G_N M^2}{25 R^4}$ would certainly change for a deeper treatment of the stellar structure. 


In this sense, it would be interesting to perform more detailed calculations considering, for instance, a neutron star composed by a solid crust surrounding a liquid core, as the one performed to study starquakes in rotation-powered pulsars \cite{franco2000}. However, at this point, we prefer to keep on the simplest model to present here order of magnitude calculations to give confidence to the suggested scenario. 

To estimate the energy released by this mechanism during the anti-glitch event, we use the classical model developed for pulsar glitches \cite{baym.pines}. Three energy contributions are considered: gravitational, coming from the global change in the stellar shape; rotational, associated to the spin frequency shift, $\Delta \nu / \nu$, and strain tensions released by the crust. This strain energy is accumulated because, despite that the magnetic field decays from $\langle B_t^{i}\rangle ^2$ to  $\langle B_t^{f} \rangle ^2$, the crystallized crust keeps its original shape of $\epsilon^-$ by increasing its internal tension, departing from equilibrium. Once the critical strain is achieved, the crust cracks and the star re-accommodates into an $\epsilon^+$ equilibrium configuration, releasing the stored energy. As the length scale, $\ell$, associated to the change in shape in our model is small $\ell/R \sim \epsilon$, we assume that magnetic field reconnection does not take place. For this reason, even though we are treating with a magnetar, we do not consider the magnetic energy release due to the displacement of magnetic field footpoints, following the standard Soft Gamma Repeater picture \cite{TD95,TD96}.

\section{Results}

\begin{figure}
 \centerline{\includegraphics[width=0.55\textwidth,angle=-90]{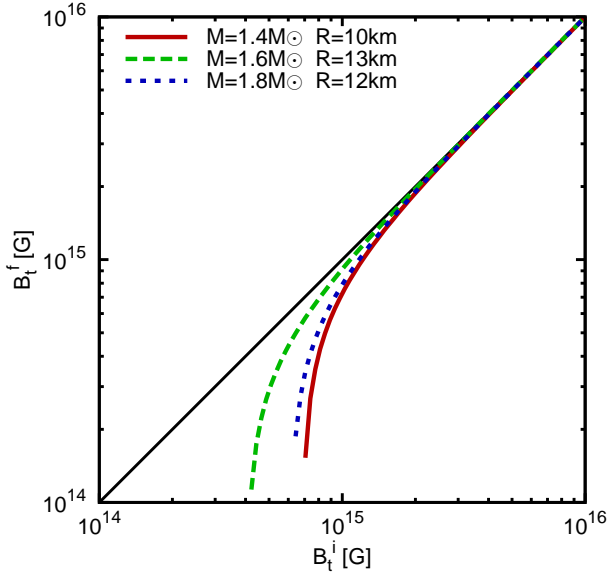}}
 \caption{Physical solutions to equation (\ref{deltanu2}) for $\Delta \nu /\nu = - 6.3 \times 10^{-7}$ as function of $\langle B_t^i\rangle$, for three different neutron star configurations (see the legend). In black we plot the identity function as a reference.}
 \label{fig1}
\end{figure}

\begin{figure}[htb]
 \centerline{\includegraphics[width=0.55\textwidth,angle=-90]{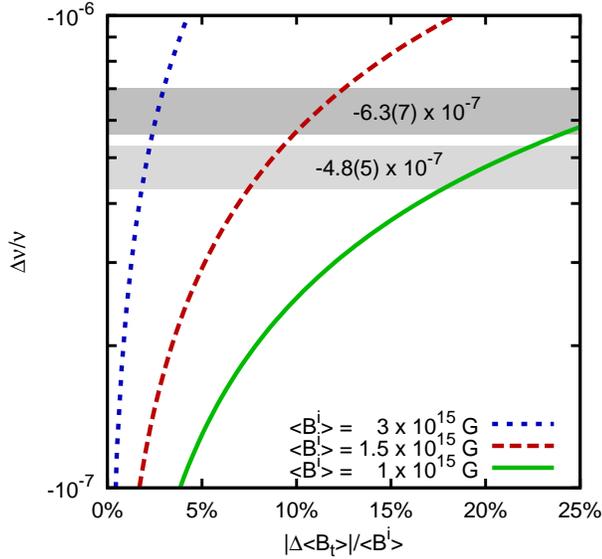}}
 \caption{$\Delta \nu / \nu$ values obtained from  equation (\ref{deltanu1}) as a function of the relative change $\left|\Delta \langle B_t\rangle\right| / \langle B_t^i\rangle$ in the mean toroidal magnetic field strength. As a reference we also show the $\Delta \nu / \nu$ values of the anti-glitch/anti-glitch pair observed in \mbox{AXP 1E 2259+586} in shaded rectangles with their corresponing error bars.}
 \label{fig2}
\end{figure}

In the framework of our model, we estimate the long-term decay in the magnetic field strength needed to account for the $\Delta \nu / \nu$ observed in the anti-glitch of magnetar \mbox{AXP 1E 2259+586}. 

As a function of the mean initial toroidal magnetic field strength, $\langle B_t^i\rangle$, and for three different neutron star configurations, we plot in Figure (\ref{fig1}), the physical solutions, $\Delta \langle B_t\rangle < 0$, to equation (\ref{deltanu2}). For this, we assume a frequency jump $\Delta \nu / \nu = -6.3 \times 10^{-7}$ equal to the first of the two events of model (ii). In black we plot the identity as a reference.

For a typical neutron star with a mean toroidal magnetic field of $\langle B_t\rangle = 2 \times 10^{15}$~G, which corresponds to a maximum value for the magnetic field strength $B_M > 10^{16}$~G \cite{reisenegger2013}, we estimate that a decay in the magnetic field of about $\sim$10\% could be responsible for the observed spin-down. These qualitative result is almost insensitive to other acceptable values for mass, radius and magnetic field strength of magnetars (see Figure \ref{fig1}). Detailed studies of the magnetic field evolution in neutron stars show that a magnetic field decay of $\sim$10\% is easily achieved after $t < 10^6$~yr for a magnetar like \mbox{AXP 1E 2259+586} \cite{vigano2013}. 

For the adopted neutron star configurations, we find that a minimum value for $\langle B_t^i\rangle $ (different in each case) is needed in order to have a solution to equation (\ref{deltanu1}) for the observed $\Delta \nu / \nu$. This critical value is, in any case, several times $10^{14}$~G, which avoids the occurrence of anti-glitches in normal pulsars, only allowing this phenomena to occur in strongly magnetized neutron stars, i.e. magnetars. This result can be used to explain why, despite many pulsars have been thoroughly monitored for several decades, no sudden spin-down event of this kind has been detected at all.

In Figure (\ref{fig2}) we present $\Delta \nu / \nu$ from equation (\ref{deltanu2}) as a function of the change in the mean toroidal magnetic field strength. The shaded rectangles represent the values of $\Delta \nu / \nu$ from model (ii) and their corresponding error bars.

We showed how our simple model can be used to explain the timing behaviour of the anti-glitch observed in magnetar \mbox{AXP 1E 2259+586}. Hence, we now focus our attention to analyze the energetics.

For a ``typical'' neutron star and assuming a normal neutron star crust, we estimate a gravitational energy release of $\sim 10^{42}$~erg, while the rotational and crustal strain contributions are several orders of magnitude smaller. Assuming a typical SGR, the magnetic energy release is $\sim 10^{41}$~erg. Hence the total energy emitted after the anti-glitch in gravitational waves, particles and electromagnetic radiation should be of the order of $10^{42}$~erg which is compatible with the observations.

\section{Conclusions}

We present here a very simple model based on an internal mechanism to explain the anti-glitch observed in magnetar \mbox{AXP 1E 2259+586}. We suggest that a long term magnetic field decay of approximately $\sim$10\%, from an initial $\langle B_t^i\rangle \sim 10^{15}$~G, would be enough to de-stabilize an originally prolate stellar configuration cracking the neutron star crust to a ``more spherical'' one. As a result, the sudden change in the moments of inertia of the star would naturally lead to a sudden spin-down, as the one observed in \cite{antiglinch}. Under this scenario, in addition, considering a ``typical'' neutron star, we estimate an energy release of $\sim$10$^{42}$~erg, which is in agreement with the emission detected by {\it Fermi} and {\it Swift} observatories in the epoch when the anti-glitch occurred. 

As a corolary, our simple model predicts that an anti-glitch as the one detected by \cite{antiglinch} in \mbox{AXP 1E 2259+586}, can only be achieved if the mean internal toroidal magnetic field of the neutron star is several times $10^{14}$~G, as in magnetars, avoiding the occurrence of anti-glitches of this amplitude in normal pulsars, which might be the reason why this event is the first of its kind to be detected.



\subsection*{Acknowledgement}

We express our thanks to the organizers of the CSQCD IV conference for providing an excellent atmosphere which was the basis for inspiring discussions with all participants.
We have greatly benefitted from this. FG and IFRS are Fellows of CONICET. IFRS acknowledges support by UNLP.

\end{document}